\documentstyle[prl,aps,twocolumn]{revtex}
\begin{document}
\title{A Critical Reexamination of the Peierls-Nabarro 
Model}

\author{Gang Lu, Nicholas Kioussis}
\address{Department of Physics,California State University Northridge,\\
 Northridge, CA 91330-8268}

\author{Vasily V. Bulatov}
\address{Lawrence Livermore National Laboratory, Livermore, CA 94551}

\author{Efthimios Kaxiras}
\address{Department of Physics, Harvard University, Cambridge, MA 02138}
\date{\today}
\address{
\begin{center}
\begin{minipage}{14cm}
\begin{abstract}
We reexamine two essential issues within the Peierls-Nabarro model which are 
critical in obtaining accurate values for the Peierls stress. 
The first issue is related to the 
sampling scheme of the misfit energy across the glide plane 
and the second one is 
the effect of relaxation on the Peierls stress. 
It is shown that most of the current 
applications of the Peierls-Nabarro model do not treat properly the two issues 
and therefore are not able to predict reliable values for the Peierls stress. 
We argue that the 
double counting scheme for the misfit energy at both sides of 
the glide plane is 
physically more reasonable, and it can predict more accurate 
values for the Peierls stress, 
especially for dislocations with equal 
spacing between 
alternating 
atomic rows. We also 
show the importance for allowing atomic relaxation when a dislocation 
traverses the Peierls barrier, which in turn lowers the Peierls stress  
for narrow dislocations by an 
order of magnitude. 
\end{abstract}
\end{minipage}
\end{center}
}
\maketitle

The Peierls stress $\sigma_p$ is the minimum applied external stress 
required to move 
a stationary dislocation irreversibly, without 
the assistance from thermal or 
quantum lattice vibrations. This fundamental quantity of a lattice 
dislocation was first estimated by Peierls \cite{1} and 
Nabarro \cite{2} using  
essentially a continuum model, which is now referred to as the Peierls-Nabarro 
 (P-N)  model.
The original P-N model has served more as a conceptual tool for
a qualitative understanding of dislocation core properties, 
rather than providing 
 quantitative estimates of these properties, due to the unrealistic 
sinusoidal force law adopted in the model. 
However, there has been renewed interest in the P-N model recently to calculate 
the Peierls energy and stress \cite{3,4,5,6,7} thanks to  
the advent of highly accurate {\it 
ab initio} total-energy methods based on the density functional theory (DFT). 
 To date, the P-N model serves as a link 
between atomistic and continuum approaches, by providing a means to 
incorporate information obtained from the atomistic ({\it ab initio} or 
empirical) calculations directly into the continuum model. This resultant 
approach  can then be applied to 
problems that neither atomistic nor conventional continuum models could handle 
separately. 

After the pioneering work of Peierls and Nabarro, several attempts 
to improve the original model have been made.  
For example, the three-dimensional dislocation profile has 
replaced the one-dimensional constrained path approximation \cite{3,8,9}
 in the 
original model. This 
is particularly important for treating the dissociation 
of a perfect dislocation 
into partials. A fully anisotropic treatment of the 
elastic interaction has 
been implemented into the model \cite{10}. 
The above extensions have  allowed 
 the study of straight dislocations with  
arbitrary orientation in an 
arbitrary glide plane and a medium of arbitrary anisotropy.  More recently, 
an averaging procedure has been proposed to account for the nonlocal nature of 
the misfit energy \cite{11}.  By discretizing the elastic energy, Bulatov and
Kaxiras \cite{2} introduced recently a 
semidiscrete variational model which takes into account the important 
degrees of freedom which participate actively in the translation of the 
dislocation over the Peierls barrier. 

In spite of the apparent importance and success of the P-N framework, 
a number of shortcomings still exist in the model.  
The classic P-N model completely 
neglects the relaxation of atoms when the dislocation center  
 moves across the 
Peierls energy barrier, e.g., the dislocation displacement profile stays
constant during the translation of the dislocation. Thus two
approximate treatments for the total energy are involved in the classic
model.
Specifically, 
the variation of the elastic energy with respect to the translation
vanishes due to the fact that the elastic energy is the functional of
the overall displacement field of the lattice which are not allowed to change 
during the
translation of the dislocation,  hence the elastic energy has no contribution 
to the 
Peierls stress. On the
other hand although the variation of the misfit energy with respect to
the dislocation translation exists due to the discrete lattice
nature, no relaxation of atoms is allowed to compute the correct (relaxed)
lattice potential resulting from the interaction between the moving 
dislocation and
the underlying lattice.
In other words, in the classic P-N model
the dislocation translates rigidly 
through the lattice. How the two approximations affect the description for 
the dislocation
core properties in still an open question, therefore 
it is one of the purposes in this work to examine the importance 
of relaxation in determining the Peierls stress and other dislocation
core properties for both narrow and intermediate
 dislocations, such as
those in Al.

The second important issue addressed in this paper is more controversial 
and subtle. It is related to the misfit energy sampling for different crystal 
lattices,
i.e. facing lattice (F-type), 
in which the atoms just above and below the glide plane
face each other, such as the simple cubic lattice or 
alternating lattices (A-type) in which the atomic sites on the two sides
of the glide plane alternatingly distribute as in a closed packing
lattice, such as the fcc lattice.
In their original work, Peierls, Nabarro and others \cite{1,2,12} 
 summed 
the misfit energy independently over the top and bottom half-crystals 
for both 
F and A-type lattices. We will refer to this scheme as
the double counting (DC) scheme. 
However, the DC scheme 
yields the variations of the misfit energy 
and 
the lattice friction which have a periodicity of $b$/2, 
in contrast with the feature of 
the dislocation barrier which must in general exhibit 
the periodicity of $b$. 
Most people believe the unexpected form of the Peierls energy barrier is 
attributable to the DC approach in the 
classic P-N model \cite{13,14}
and since then 
almost all numerical and analytical work employ exclusively a 
single counting 
(SC) scheme \cite{4,5,6,7}, e.g., the scheme in 
which one sums the misfit energy only on one side of 
the glide plane as a function of the disregistry 
(relative displacement) between 
pairs of atomic rows. The argument for using the SC approach
is based on the assumption of the nearest-neighbor 
 interaction and on the fact that this scheme gives the correct 
periodicity of $b$. However, as Wang pointed out recently, the undesired 
periodicity is actually due to an erroneous representation 
of the atomic positions in the original model 
rather than the DC scheme itself \cite{15,16}.  
In fact, the DC scheme $can$ give the correct periodicity for the both 
F and 
A-type lattices after the error of atomic positions is corrected. 
On the other hand the accuracy of  
the nearest-neighbor interaction approximation in the SC scheme for the
A-type lattice is
still questionable. 
Thus the second aim of this paper is to examine the 
soundness of the two schemes 
and compare for the first time the misfit energy and the Peierls 
stress calculated from these schemes. 
Finally we present a simple analytical expression
for the Peierls stress derived for the A-type lattice 
based on the DC scheme.

In order to study these subtlies within the P-N model, we 
employ the semidiscrete 
variational P-N model, because of its accuracy and expedition
\cite{3,8}. 
For the study of dislocations in Al (intermediate core width)
the $\gamma$
surface is obtained by means of {\it ab initio} pseudopotential total-energy
calculations \cite{8}.
For the case of extremely narrow dislocations, we will
resort to a sinusoidal force law for calculating the misfit energy and the 
Peierls stress.
Within the semidiscrete variational approach 
the equilibrium structure of a dislocation is obtained by minimizing 
 the dislocation energy density functional 
\begin{eqnarray}
U_{disl}&=& 
\sum_{i,j}\frac{1}{2}\chi_{ij}[K_e(\rho_i^{(1)}\rho_j^{(1)} + 
\rho_i^{(2)}\rho_j^{(2)}) + K_s\rho_i^{(3)}\rho_j^{(3)}] \nonumber \\  
& & 
+ \sum_i\Delta x \gamma_3(\vec{\delta_i})
- \sum_{i,l}\frac{x_i^2-x_{i-1}^2}{2}(\rho_i^{(l)}\tau_i^{(l)})  
\label{ene_fun}
\end{eqnarray}
with respect to the dislocation densities or displacement vectors. 
Here $\rho_i^{(1)}$, $\rho_i^{(2)}$ and $\rho_i^{(3)}$ 
are the edge, vertical and 
screw components of the general interplanar displacement density 
at the $i$-th nodal point, and $\gamma_3(\vec{\delta}_i)$ is the 
three-dimensional misfit potential.
The corresponding applied stress components interacting 
with the $\rho_i^{(1)}$, $\rho_i^{(2)}$ and 
$\rho_i^{(3)}$, are 
 $\tau^{(1)} = 
\sigma_{21}$, $\tau^{(2)} = \sigma_{22}$ and 
$\tau^{(3)} = \sigma_{23}$,
 respectively. $K_e$ and $K_s$ are the edge and screw 
pre-logarithmic energy factors. 
The dislocation density at the $i$-th nodal point is 
$\rho_i = (\delta_i - \delta_{i-1})/(x_i - x_{i-1})$,
where $\delta_i$ and $x_i$ are the displacement vector and 
the coordinate of the 
$i$-th nodal point (atomic row), respectively,
 and
$\chi_{ij} = \frac{3}{2}\phi_{i,i-1}\phi_{j,j-1} + 
\psi_{i-1,j-1} + \psi_{i,j} 
- \psi_{i,j-1} - \psi_{j,i-1}$, with $\phi_{i,j} =  
x_i - x_j$, and $\psi_{i,j} = 
\frac{1}{2}\phi_{i,j}^2\ln\vert\phi_{i,j}\vert$.
Because the displacement vector $\vec{\delta}(x_i)$ 
is allowed to change during the 
process of dislocation translation, the Peierls energy barrier can be 
significantly lowered compared to its corresponding value from a rigid 
translation. The response of the dislocation 
to an applied stress is achieved by the minimization of the 
energy functional with 
respect to $\rho_i$ at a given value of the applied stress, 
$\tau_i^{(l)}$. The 
Peierls stress is then obtained by evaluating the 
critical value of the applied 
stress $\tau_i^{(l)}$ at which the dislocation energy functional 
fails to be 
minimized with respect to $\rho_i$ through standard conjugate gradient 
techniques. This approach of calculating the Peierls 
stress is more accurate and 
physical transparent because it captures the nature of 
the Peierls stress as the 
stress at which the displacement field of the dislocation undergoes a 
discontinuous transition. 

In Table I, we compare the values of the misfit energy and the 
Peierls stress for various dislocations in Al, 
calculated from the DC and SC schemes.
 The {\it ab initio} determined $\gamma$
surface has been implemented into the semidiscrete model 
to obtain accurate values for the 
energetics and the Peierls stress.
 All 
dislocations considered here are perfect fcc dislocations in (111)
plane
 but with different 
orientations, e.g., different angles between the dislocation lines 
and the 
Burgers vectors.  
For all dislocations in Al, the 
atomic rows from the two sides of the glide plane are situated 
alternatingly from each other. 
 Furthermore it is important to note 
that, with the exception of the $30^\circ$ and $90^\circ$ dislocations,  
the spacing between alternating atomic rows
for all the other dislocations in an fcc lattice is not even.
We will refer to the $30^\circ$ and $90^\circ$ dislocations as
evenly-spaced dislocations.
Two conclusions can be drawn from the comparison in 
Table I. First the SC scheme always underestimates the misfit energy and 
overestimates the Peierls stress compared to the DC scheme. 
Second the agreement for the Peierls stress  between 
the two schemes becomes particularly worse for the 
evenly-spaced ($30^\circ$ and $90^\circ$) dislocations.
Namely, while the SC scheme 
underestimates the misfit energy by about 15\%, it 
overestimates the Peierls stress relative to its value from the DC scheme
by orders of magnitude .
Thus the even versus uneven spacing between atomic rows turns out to be 
an important issue regarding the Peierls stress, and has 
not been addressed in previous studies. 

In order to establish which 
scheme is superior over the other and to 
understand the different results received from two sampling schemes, 
next we take a closer look at the two schemes for the A-type lattice
(the two schemes are essentially the same for the F-type lattice).
In the SC scheme, the misfit energy is sampled as a function of the 
relative displacement 
 between pairs of atomic planes across the glide plane. 
Therefore, the SC 
model considers only the nearest-neighbor interaction 
across the glide plane, e.g., 
the local bonding distortion between the pair of atomic rows. Although this 
model seems to be applicable to covalently-bonded systems, where the bonding 
across the glide plane is highly localized, 
it fails to describe metallic 
systems, which have more delocalized electronic states. Along the same line of 
thinking, one can understand why the SC model works particularly worse 
for the evenly-spaced (30$^\circ$ and 90$^\circ$)
 dislocations in which the atomic rows below (above) the glide plane are 
located right in the
middle of the atomic rows above (below) the glide plane. In this 
case it is most ambiguous to define a 
local atomic pair, and the second nearest-neighbor interaction is as important 
as the first nearest-neighbor interaction. 
The neglect of higher nearest-neighbors interactions in the SC scheme  
gives rise a lower misfit energy and a much higher 
Peierls stress. 
On the other hand the DC scheme treats the misfit energy from the 
top and bottom crystal as separate entities, 
and the misfit energy is defined as a 
function of the displacement between the original position of an atomic row and 
its final position after the introduction of a dislocation. 
Consequently the misfit energy
can be summed up independently over the top and bottom half-crystals. 
Two advantages 
immediately emerge from the DC model: (1) Higher-order interactions are 
included naturally. In fact   
the misfit energy is exact, by including the 
contributions from all atomic rows across the glide plane. In other
words the 
misfit energy is the result from the overall charge density 
redistribution due to 
the displacements of all atoms on both sides of the glide plane, rather than 
the local bonding distortion associated with the SC scheme. 
(2) The fact that the DC approach samples the misfit energy 
over twice as many atomic rows as in the SC method, it reduces the error due to 
the local displacement gradient approximation by a factor of two \cite{17,18}.
This improvement is particularly important for 
describing narrow core dislocations  
where the displacement gradients are relatively larger.
 
Next we present for the first time a simple and rigorous analytical formula 
of the Peierls stress for  
dislocations in A-type lattice based on the DC scheme. 
Although this formula is 
derived using the sinusoidal approximation for the restoring force, 
it will provide insight on the effect of atomic spacing on the 
Peierls stress
and will allow a qualitative understanding of the results in Table I. 
Following the SC
treatment of Joos and Duesbery \cite{4}, 
 the total misfit energy within the DC scheme can be written as 
the sum of misfit energy contributions from the two half-crystals
\begin{eqnarray}
 W (u)&=&  \sum_{n=-\infty}^{+\infty} \frac{a_1 + a_2}{2}\{\gamma [\delta (n(a_1 + 
a_2) - u)] \nonumber \\
& & + \gamma [\delta (n(a_1 + a_2) + a_1- u)]\},
\label{en2}
\end{eqnarray}
where $a_1$ and $a_2$ are the spacings for the   
atomic rows alternating across the glide plane, which can be different 
in general.
As alluded earlier,  $\delta$ is the displacement field of 
atomic rows 
relative their original positions, $\gamma$ is the misfit 
potential, and u is the dislocation translation distance. 
Introducing $a = a_1 + a_2$ and $u\prime = -(a_1-u) = u - a_1$, we have
\begin{eqnarray}
W (u, u\prime)& =& \sum_{n=-\infty}^{+\infty} \frac{a}{2} \{\gamma [ \delta (na 
- u)] + \gamma [\delta (na - u\prime)]\}\nonumber \\
&=& W_1 (u) + W_2 (u\prime).
\label{en3}
\end{eqnarray}
Assuming a sinusoidal force law  for the restoring force 
$F(\delta(x))=\tau_{max} 
\sin(2\pi\delta(x))/b$, 
gives 
a misfit energy functional 
$\gamma(\delta(x))$ of the form, 
\begin{equation}
 \gamma [\delta (x)] = \frac{\tau_{max} b}{2 \pi} (1 - \cos \frac{2 \pi \delta 
(x)}{b}).
\end{equation}
The solution of the P-N integro-differential equation gives
 $\delta (x) = b/\pi\arctan(x/\zeta) + b/2$, 
where $b$ is the Burgers vector of the dislocation and $\zeta$ is the 
half-width. Substituting the expressions for $\gamma(x)$ and 
$\delta(x)$ in Eqn. 
(3) and using the Poisson's summation formula, the  
misfit energy reduces to 
\begin{eqnarray}
W(y_1, y_2)& =& \frac{K b^2 \sinh 2\pi\Gamma}{8 \pi} (\frac{1}{\cosh 2\pi\Gamma - 
\cos 2\pi y_1}\nonumber \\
& &
+ \frac{1}{\cosh 2\pi\Gamma - \cos 2\pi y_2}), 
\label{en4}
\end{eqnarray}
where $y_1 = u/a$, $y_2 = u\prime/a = (u-a_1)/a = 
y_1 - a_1/a$, and $\Gamma = \zeta/a$.  $K$ is a constant depending
on the elastic 
properties of the dislocations. 
It should be pointed out that the Eqn. (5)  is 
valid only for non-evenly spaced dislocations due to the requirement 
imposed by the 
Poisson's summation formula. 
It is straightforward however, to calculate the Peierls 
stress for evenly-spaced dislocations using a similar approach. To get the 
Peierls stress one needs to calculate the stress associated with 
the misfit energy 
variation. 
\begin{eqnarray}
\sigma(y_1,y_2)&  =& \frac{1}{b} \frac{d W}{d u} 
 =  -\frac{K b \sinh2\pi\Gamma}{4a} [\frac{\sin2 \pi y_1}{(\cosh 2 \pi \Gamma - 
\cos 2 \pi y_1)^2} \nonumber \\& & + \frac{\sin2 \pi y_2}{(\cosh 2 \pi \Gamma - \cos 2 \pi 
y_2)^2}].
\label{en5}
\end{eqnarray}
Letting $t=a_1/a$ and $t \not= \frac{1}{2}$, 
\begin{eqnarray}
\sigma (y_1, t)& =& -\frac{Kb\sinh 2\pi \Gamma}{4a} [\frac{\sin2\pi 
y_1}{(\cosh2\pi\Gamma - \cos2\pi y_1)^2}\nonumber \\& &
+ \frac{\sin2\pi (y_1 - 
t)}{(\cosh2\pi\Gamma - \cos2\pi (y_1 - t))^2}]
\label{en6}
\end{eqnarray}
For all dislocations studied in Al, we find $\Gamma = \frac{\zeta}{a} > 1$ 
 and this should hold also for not very narrow 
dislocations. In this limit, Eqn. (7) reduces to
\begin{equation}
\sigma (y_1, t) = -\frac{Kb\cos \pi t}{2a\cosh2\pi \Gamma} \sin(2\pi y_1 - 
\pi t).
\end{equation}
For a given $t$, the maximum value of $\sigma (y_1, t)$ 
corresponds  to the Peierls stress 
\begin{equation}
\sigma_p = \frac{Kb}{a}e^{-2\pi \zeta/a} |\cos(\pi t)|
\end{equation}
For comparison we present below also the formula derived by Joos and Duesbery 
using the SC model [Eqn. (24) in Ref. 4]  
\begin{equation}
\sigma_p = \frac{Kb}{a\prime}e^{-2\pi \zeta/a\prime}.
\end{equation}
Here $a\prime = a$ . Two results are evident from 
the simple analytical expression derived above. First, Eqn. (8) 
shows that the DC scheme gives the 
correct periodicity $a$. Second, the value of the Peierls stress 
calculated from the DC scheme is always 
smaller than that from the SC scheme. Both these analytical results 
are consistent with our 
numerical results listed in Table 1. 
For the non-evenly spaced dislocations in the fcc lattice, 
t=2/3, 
thus 
we find that the SC scheme predicts a Peierls stress which is twice 
as large as that obtained 
from the DC scheme using the sinusoidal force law. 
For the evenly-spaced dislocations, it can be 
easily shown that
\begin{equation}
\sigma_p = \frac{2Kb}{a}e^{-4\pi\zeta/a}, 
\end{equation}
and the ratio $m(\zeta)$ of the Peierls stress from the SC ($\sigma_p^{SC}$) 
to the DC ($\sigma_p^{DC}$) scheme is
\begin{equation}
m(\zeta) = \frac{\sigma_p^{SC}}{\sigma_p^{DC}} = 
\frac{1}{2}e^{2\pi \zeta/a}.
\end{equation}
This equation shows that 
 for the evenly spaced
dislocations (30$^\circ$ and 90$^\circ$), 
the SC scheme predicts a much higher Peierls
stress than the DC scheme.
It also demonstrates that $m(\zeta)$ increases exponentially 
with $\zeta$, yielding a larger $m(\zeta)$ ratio 
for the 90$^\circ$ dislocation compared to that for the 30$^\circ$,
due to the larger half-width of the 90$^\circ$ dislocation \cite{8}. 
Although this simple formula gives the correct qualitative trend for
$\sigma_p$, 
it is rather limited in predicting reliable Peierls stress for real materials
 because of the sinusoidal force law employed in its 
derivation. For example one problem associated with the  
sinusoidal force law is that it yields no dissociation 
into partials, which can 
result in several orders of magnitude difference in the $\sigma_p$ if the 
dissociation is allowed. It needs to be emphasized
that in order to test the reliability of
different models,  
one needs to compare to direct atomistic calculations.
The semidiscrete model with the DC scheme has been found  
 to be able to predict reliable values for the $\sigma_p$ for both
covalent and metallic systems, e.g., in
 good agreement
with the direct atomistic simulation results using the same interatomic 
potentials \cite{3,8}. 

We next examine the effect of atomic relaxation on the Peierls stress 
for different 
types of dislocations in Al. 
In Table II, we list the $\sigma_p$ 
 for different dislocations 
using three different approaches.
All the three methods are based on the DC scheme. The first (SD+DFT) 
and second (SD+sin) methods employ the semidiscrete model but with 
different restoring forces. Namely the SD+DFT method 
uses a restoring 
force obtained from density functional calculations, 
whereas the second method (SD+sin) uses a sinusoidal force law,
but with the $\tau_{max}$ in Eqn.(4) adjusted to give the same value 
for the dislocation half-width ($\zeta$) as in the SD+DFT calculations.
The third method uses the analytical expressions (Eqns. (9) and (11)) 
for the $\sigma_p$ using
the same $\zeta$ as for the first two methods.
Therefore the comparison between the (SD+DFT) and 
(SD+sin) methods provides information for the effect of sinusoidal
approximation 
on $\sigma_p$ (both schemes include relaxation), 
while the comparison between the second and third methods illustrates 
the effect of atomic
relaxation. It is interesting to note that the relaxation effect is small  
for all dislocations in Al,
whereas the approximation of the sinusoidal restoring force is not good, 
as we addressed earlier.

Recently Schoeck claimed that the relaxation effect 
on the Peierls stress can be canceled out by the 
opposite contributions from the misfit energy and the 
elastic energy \cite{11}. 
Thus we need to clarify the cause of the negligible relaxation effect in Al, 
i.e., whether it is due to the small Peierls energy barrier or 
the cancellation effect. 
We have carried out two sets of calculations based on the DC 
scheme and a sinusoidal 
 force law, but   
$\tau_{max}$ is chosen large enough 
(1.2 ev/\AA$^3$) to produce narrow dislocations ($\zeta$ less 
than 0.5 \AA~) in the Al lattice.
The first set of calculations employs the semidiscrete model 
which allows relaxation while the second set is based 
on the analytical formula $\sigma_p =
\frac{3\sqrt{3}}{2} \tau^2_{max} a/Kb$ 
derived for narrow dislocations \cite{4} (no relaxation is included). 
Note for very narrow dislocation the misfit energy is localized 
within one lattice spacing and hence the details of the misfit
energy sampling (DC vs SC) is not relevant.
Listed in Table III are the results of the $\sigma_p$ for 
different dislocations from these two sets of calculations.
One can see that the relaxation effect is
significant (more than one order of magnitude)
and no canceling effect is observed. 

In summary we have examined two essential issues 
within the Peierls-Nabarro model 
which are critical to obtain accurate values for the Peierls stress. 
The first issue concerns the 
sampling scheme for the misfit energy. We have shown that the DC scheme is more 
appropriate for determining the Peierls stress, especially for evenly-spaced 
dislocations, 
although most work in the literature has done otherwise. 
An analytical formula is 
derived to help understanding the  
results received from the numerical calculations. 
The second issue addressed in this work is the effect of relaxation 
 for various dislocations.  We have demonstrated the importance of 
relaxation in obtaining reliable Peierls stress for narrow dislocations. 
Comparison of the Peierls stress using {\it ab initio}  
and the sinusoidal restoring force 
 illustrates the failure of the sinusoidal approximation.

The work at the California State University Northridge was supported by the US 
Army Research Office through the grant No. DAAG55-97-0093.

\newpage
\begin{table}
\caption{Comparison of the misfit energy (ev/\AA) and Peierls stress 
(mev/\AA$^3$) employing  the DC and SC schemes. 
The semidiscrete model with the {\it 
ab initio} $\gamma$ surface is used for the calculations. }
\begin{tabular}{cccc}
& & DC & SC \\ \hline
0$^\circ$ & misfit energy & 0.0994 & 0.0908 \\
          & Peierls stress & 1.60   & 3.90 \\ \hline
25.3$^\circ$ & misfit energy & 0.1229 & 0.1173 \\
             & Peierls stress & 0.04 & 0.14 \\ \hline
30$^\circ$   & misfit energy & 0.1221 & 0.1024 \\
             & Peierls stress & 0.33  & 3.40   \\ \hline
44.7$^\circ$  & misfit energy & 0.1370 & 0.1280  \\
              & Peierls stress & 0.37 & 0.84 \\ \hline
60$^\circ$  & misfit energy  & 0.1521 & 0.1356 \\
            & Peierls stress & 0.61   & 1.30 \\ \hline
85.3$^\circ$ & misfit energy & 0.1706 & 0.1630 \\
             & Peierls stress & 0.02  & 0.07 \\ \hline
90$^\circ$  & misfit energy & 0.1688 & 0.1470 \\
            & Peierls stress & 0.02  & 2.20 \\    
\end{tabular}
\end{table}                                                         

\begin{table}
\caption{The Peierls stress (mev/\AA$^3$) for dislocations 
in Al determined through 
three different approaches. 
The DC scheme is
 used in the semidiscrete model 
(SD) and the analytical formula calculations. For the same dislocation all the 
three methods use the same core width in order to give a fair comparison.}
\begin{tabular}{cccc}
             & SD+DFT  & SD+sin   & Formula \\ \hline
0$^\circ$    & 1.60    & 0.65     & 0.99  \\
30$^\circ$   & 0.33    & 0.02     & 0.04  \\
44.7$^\circ$ & 0.37    & 0.04     & 0.06 \\
55.3$^\circ$ & 0.53    & 0.07     & 0.12 \\
60$^\circ$   & 0.61    & 0.09     & 0.14 \\
66.6$^\circ$   & 0.44    & 0.03     & 0.05 \\
\end{tabular}
\end{table}

\begin{table}
\caption{The Peierls stress (ev/\AA) for dislocations in Al lattice 
calculated from 
the semidiscrete model (SD) and the analytical formula derived only for narrow 
dislocations (see text). The same sinusoidal restoring force with the magnitude 
$\tau_{max}$ = 1.2 ev/\AA$^3$ is used for both methods. The core widths for all the 
dislocations studied here are less than 0.5 \AA.}
\begin{tabular}{ccc}
            & SD+sin   & Formula \\ \hline
 0$^\circ$  & 1.2      & 18.0   \\
 30$^\circ$  & 0.9     & 18.4  \\
 60$^\circ$  & 1.1     & 12.9  \\
 90$^\circ$  & 0.7     & 13.7  \\
\end{tabular}
\end{table}

\newpage
\begin{references}
\bibitem{1} R. E. Peierls, Proc. Phys. Soc. London {\bf52}, 34 (1940).
\bibitem{2} F. R. N. Nabarro, Proc. Phys. Soc. London {\bf59}, 256 (1947).
\bibitem{3} V. Bulatov and E. Kaxiras, Phys. Rev. Lett. {\bf78}, 4221 (1997).
\bibitem{4} B. Joos and M. Duesbery, Phys. Rev. Lett. {\bf78}, 266 (1997).
\bibitem{5} B. Joos, Q. Ren and M. Duesbery, Phys. Rev. B {\bf50}, 5980 (1994).
\bibitem{6} Y. Juan and E. Kaxiras, Philos. Mag. A {\bf74}, 1367 (1996).
\bibitem{7} J. Hartford, B. von Sydow, G. Wahnstrom and B. I. Lundqvist, Phys. 
Rev. B {\bf58}, 2487 (1998).
\bibitem{8} G. Lu, N. Kioussis, V. Bulatov and E. Kaxiras (submitted to
Philos. Mag. A).
\bibitem{9} G. Schoeck, Philos. Mag. Lett. {\bf77}, 141 (1998).
\bibitem{10} G. Schoeck, Philos. Mag. A {\bf69}, 1085 (1994).
\bibitem{11} G. Schoeck, Phys. Rev. Lett. {\bf82}, 2310 (1999).
\bibitem{12} H. Huntington, Proc. Phys. Soc. London {\bf68}, 1048 (1955).
\bibitem{13} J. P. Hirth and J. Lothe, {\it Theory of Dislocations} (Wiley, New 
York, 1982), 2nd ed., p.237.
\bibitem{14} J. W. Christian and V. Vitek, Rep. Prog. Phys. {\bf33}, 307 (1970).
\bibitem{15} J. N. Wang, Acta Mater. {\bf44}, 1541 (1996).
\bibitem{16} J. N. Wang, Mater. Sci. Eng. A {\bf206}, 259 (1996).
\bibitem{17} R. Miller and R. Phillips, Philos. Mag. A {\bf73}, 803 (1996).
\bibitem{18} R. Miller, R. Phillips, G. Beltz and M. Ortiz, J. Mech. Phys. Solids 
{\bf46}, 1845 (1998). 
\end{references}
\end{document}